\documentclass[12pt]{article}

\usepackage[a4paper]{geometry}

\usepackage[utf8]{inputenc}
\usepackage[english, german]{babel}
\usepackage{setspace}

\usepackage[T1]{fontenc}
\usepackage{float}
\usepackage{graphicx}
\usepackage{wrapfig}
\usepackage{fmtcount}
\usepackage{rotating}
\usepackage[table]{xcolor}
\usepackage{tabularx}

\usepackage[varg]{txfonts}

\begin{document}

\selectlanguage{german}

\begin{figure*}[h]
	\begin{center}
	\includegraphics[width=0.6\textwidth]{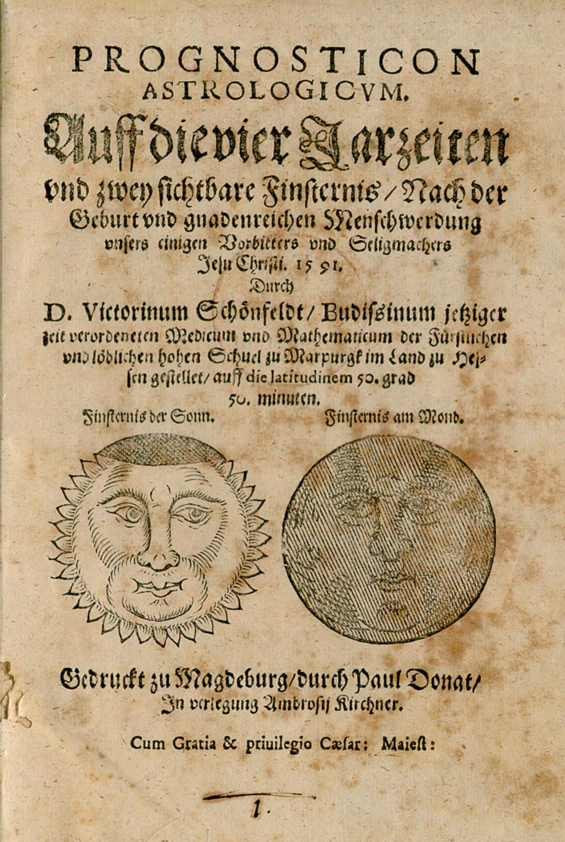}
		\label{prog_1591}

	\vspace{1cm}
			
	Titelblatt des Prognosticons für das Jahr 1591 von Victorinus Schönfeldt \cite{Schoenfeldt1590a}. Scan: Staatsbibliothek Berlin.
	\end{center}

\end{figure*}

\pagebreak

%\title{
{\Large

\noindent	
Victorinus Schönfeldt (1533 -- 1591) \\und sein ''Prognosticon Astrologicum`` 
}
%}

{\em \noindent Andreas Schrimpf (Marburg) }

%\abstract{%
%\begin{abstract}

\vspace{5mm}

Kalender und Prognostica gehörten nach der Erfindung der Druckkunst im 15. Jahrhundert zu den frühen schon recht weit verbreiteten Druckwerken. In der Landessprache, nicht in Latein, der damaligen Sprache der Wissenschaft, wurden der Lauf der Sonne, des Mondes und der Planeten dargestellt und zur Vorhersage von Wetter, Wachstum auf dem Acker und im Garten, Krankheiten und größerem Unglück benutzt. Auf diese Weise wurden astronomische Zusammenhänge – wenn auch mit astrologischer Interpretation – jedem, der lesen konnte, zugängig gemacht. 

Bedeutend für astronomische Berechnungen der Positionen der Körper unseres Sonnensystems war im Jahr 1543 die Veröffentlichung des heliozentrischen Modells von Copernicus. Victorinus Schönfeldt begann 1551 sein Studium in Wittenberg und lernte dort die Wittenbergsche Interpretation des copernicanischen Weltbildes kennen. Ab 1557 war er Professor für Mathematik an der Universität Marburg und ab 1566 übernahm Schönfeldt zusätzlich eine Professur in Medizin. Über seine Wittenberger Lehrer und Freunde erhielt er Zugang zum Hofe des Landgrafen in Marburg und wurde ein enger Freund der Familie. Er war Leibarzt von Wilhelm IV., Landgraf von Hessen-Kassel ab 1567, und wissenschaftlicher Berater bei dessen astronomischen Studien.

Er begann vermutlich direkt nach seiner Berufung an die Universität Marburg mit der Veröffentlichung seines ''\emph{Prognosticon Astrologicum}'', einer Reihe, die er bis an sein Lebensende jeweils für das folgende Jahr verfasste. Für die Berechnungen verwendete er das copernicanische Weltbild und verglich in seinen Prognostica mehrfach die Rechnungen mit denen
nach neueren ptolemaischen Ephemeriden und zum Schluss auch mit Rechnungen nach Tycho Brahes Modell.

\vspace{5mm}

Annual astrological calendars, practica and prognostications became widespread publications in Europe after the invention of printing presses in the 15th century. Using the national language instead of Latin, the language of the scientists, the motion of sun, moon and planets were explained and used to foretell weather, growth of fruit, diseases,  war and misfortune. By this means astronomical knowledge became accessible to everyone capable of reading.

A milestone in the context of position calculations of bodies of our solar system was the publication of the Copernican system in 1543. When Victorinus Schönfeldt started his studies at Wittenberg University he was educated in the so called ''Wittenberg Interpretation'' of Copernicus. In 1557 Schönfeldt became professor of mathematics at Marburg University and in 1566 he additionally was given a professorship of medicine With the help of his teachers and mentors in Wittenberg Schönfeldt was introduced to the court of the landgrave of Hessen in Marburg and became a close friend to the noble family. To  Wilhelm IV., landgrave of Hessen-Kassel from 1567, he served not only as one of the personal physicians but as a scientific counselor in the landgraves' astronomical studies.

Presumably right after being appointed a professor of mathematics Schönfeldt began to write his ''\emph{Prognosticon Astrologicum}'', a series of annual books, that he continued until his death. He made use the Copernican system for his calculations, comparing them occasionally with calculations using new Ptolemaic based ephemerides and in his last almanacs with calculations based on Tycho Brahes' model.

%\end{abstract}

%\maketitle

\section{Biografie}

Es gibt nur wenige Dokumente über den Lebensweg von Victorinus Schönfeldt. Sein akademischer Weg ist in den Unterlagen der Universitäten festgehalten und in Gelehrtenbüchern aufgelistet. Das älteste dem Autor bekannte Gelehrtenbuch stammt aus dem Jahre 1802 \cite{Strieder1802}. Durch einen glücklichen Zufall wurden im 19. Jahrhundert Schreibkalender aus den Jahren 1555 -- 1563 entdeckt, die mit Notizen von Victorinus Schönfeldt versehen sind. Daraus hat Reinhold Bechstein eine Biografie einschließlich einiger Einsichten von Victorinus Schönfeldt in das akademische Leben dieser Jahre erstellt und herausgegeben \cite{Bechstein1875}.

\begin{figure}[H]
	\begin{center}
	\includegraphics[width=0.6\linewidth]{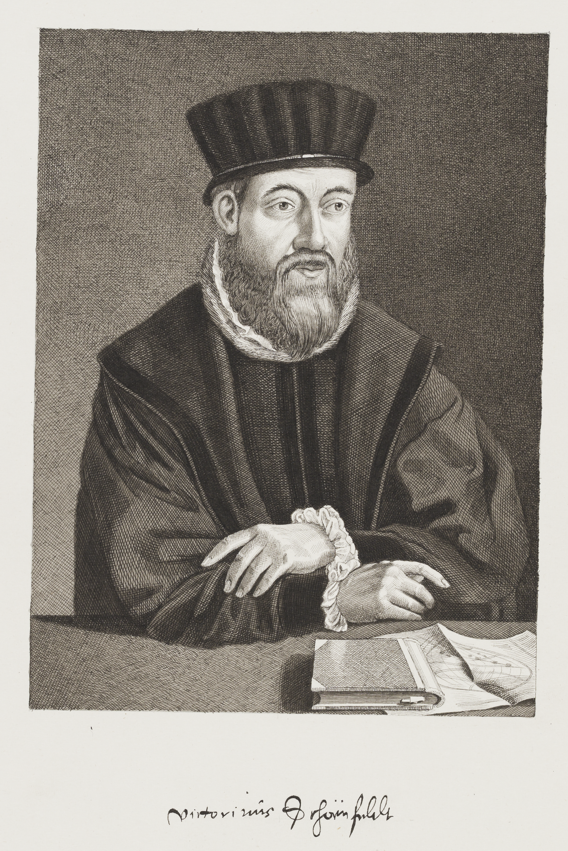}
	\caption{Victorinus Schönfeldt (1533 - 1591) im Jahr 1564. Federzeichnung in \cite{Justi1888}, Bild: Bildarchiv Foto Marburg, Fotograf: Thomas Scheidt}
	\label{fig_schoenfeldt_portrait}
	\end{center}
\end{figure}

Victorinus Schönfeldt (Abb. \ref{fig_schoenfeldt_portrait}) wurde 1533 in der Stadt Bauzen geboren und verstarb im Alter von 58 Jahren am 13. Juni 1591 in Marburg. In der Literatur wird sein Geburtsdatum oft mit 1525 angegeben (z.B.\cite{Strieder1802}), aber einen Beleg dafür gibt es nicht. Schönfeldts Sohn Burckhard gibt im Vorwort zum letzten Prognosticon seines Vaters für das Jahr 1592 dessen Alter bei seinem Tod in 1591 mit 58 Jahren an \cite{Schoenfeldt1591a}, so dass als Geburtsjahr eher 1533 anzunehmen ist.

Im April 1551 immatrikulierte er sich an der Universität Wittenberg und lernte dort bald Caspar Peucer (1525 -- 1602) und Philipp Melanchton (1497 -- 1560) kennen \cite{Bechstein1875}. Peucer, der hauptsächliche Lehrer von Schönfeldt, stammte ebenfalls aus Bauzen. Er nahm sich Schönfeldts an und kümmerte sich um sein Fortkommen; zwischen beiden entwickelte sich eine lebenslange Freundschaft. Peucer war mit der zweiten Tochter Melanchtons verheiratet, so dass Schönfeldt durch die Bekanntschaft zu Peucer auch gesellschaftlichen Umgang mit Melanchton erhielt \cite{Bechstein1875}. Schönfeldt studierte in der Artistenfakultät, schloss sich aber auch der medizinischen Fakultät an. Im August 1554 wurde er zum Baccalaureus und am 16. Februar 1557 zum Magister Artium promoviert \cite{Koestlin1891}.

Auf Empfehlung von Philipp Melanchton erhielt Schönfeldt eine Professur in Mathematik an der Universität Marburg, die er am 29. April 1557 antrat. Als sich im Jahr 1560 an der Universität eine Vakanz in Medizin auftat, übernahm nach einigen Verhandlungen 1562 Schönfeldt diese Professor vertretungsweise und arbeitete dann an seiner Promotion in Medizin. Im Januar 1556 legte er seine Disputation ab. Am 25. Mai 1556 wurde er mit der Übergabe seiner Doktorinsignien zum Professor für Medizin ernannt \cite{Caesar1877} und erhielt ein stattliches doppeltes Professorengehalt. 

Neben seiner Tätigkeit an der Universität übernahm Schönfeldt auch die Aufgabe des Leibarztes der Landgrafen von Hessen, belegt ist diese ab 1565 \cite{Saloch2006}. Eine Vergütung ist dafür nicht nachzuweisen; man geht davon aus, dass er im Krankheitsfall zu Hof gerufen wurde und dies durch das doppelte Professorengehalt als abgedeckt galt. Eine zusätzliche lukrative Einnahmequelle hatte Schönfeldt durch seine Privatpraxis, in der er hochgestellte Personen außerhalb der landgräflichen Familie kurierte \cite{Schmitz1978}.

Nachdem Schönfeldt die Medizinprofessur vertretungsweise übernommen hatte, zögerte er nicht lange, sich zu vermählen. Er heiratete am 20. Oktober 1592 Kunigunde Nordeck, die Tochter einer Ratsfamilie aus Kassel \cite{Bechstein1875}. Aus dieser Ehe gingen mehrere Kinder hervor. Der Sohn Burckhard Schönfeldt studierte in Marburg Medizin und wurde später Arzt in Fulda. 
Im Nachruf für seinen Vater schrieb Burckhard 
von der Trauer seiner Mutter, ''\emph{seinem Geschwister}'' und seiner eigenen über den Verlust \cite{Schoenfeldt1591a}. Es muss also mindestens noch ein Geschwister gegeben haben. Burckhard selber begann einige Jahre nach dem Ableben seines Vaters ebenfalls mit dem Verfassen von Prognostica \cite{Herbst2014}.

\section{Beziehung zu den Landgrafen von Hessen}

Die Universität Marburg wurde 1527 von Landgraf Philipp I. von Hessen (1504 -- 1567) als erste protestantische Hochschule der Welt gegründet.
Philipp stand in engem Kontakt zu den Reformatoren in Wittenberg. So lud er 1529 zum sogenannten Marburger Religionsgespräch ein, bei dem unter anderem auch Philipp Melanchton, einer der Lehrer Schönfeldts, zugegen war.

Aus Landgraf Philipps erster Ehe gingen vier Söhne, Wilhelm IV.\ von Hessen-Kassel (1532 -- 1592), Ludwig IV.\ von Hessen-Marburg (1537 -- 1604), Philipp II.\ von Hessen-Rheinfels (1541 -- 1583) und Georg I.\ von Hessen-Darmstadt (1547 -- 1596), hervor. Das Erbe teilte Philipp 1558 unter diesen vier Söhnen auf, wobei Wilhelm IV.\ mit Hessen-Kassel etwa die Hälfte des gesamten Gebietes zugesprochen bekam \cite{Wilhelm1558}.

Wilhelm IV.\ war sehr an den Naturwissenschaften interessiert, besonders an der Astronomie. Er ließ 1560 die erste neuzeitliche Sternwarte Europas im Kasseler Schloss errichten \cite{Mackensen1979}. Die hauptsächliche Aufgabe dieser Sternwarte war die Erstellung eines neuen Sternenkatalogs mit deutlich genaueren Positionsangaben. Außerdem wurden Untersuchungen der Parallaxe von Kometen vorgenommen und Fragen der Kalenderreform behandelt \cite{Hamel2002}.

Caspar Peucer erfuhr von der offenen Stelle eines Mathematikprofessors an der Marburger Universität, als Schönfeldt sein Examen ablegte.
So erhielt Schönfeldt mehrere Empfehlungsschreiben für seine Bewerbung, darunter eines von Philipp Melanchton an den jungen Landgrafen Wilhelm. Mit diesem Empfehlungsschreiben stellte sich Schönfeldt an der Universität Marburg vor und wurde daraufhin zu einer Audienz bei Wilhelm IV.\ geladen \cite{Bechstein1875}. In seiner Kometenschrift, die Schönfeldt Wilhelm IV.\ widmet, erwähnt er dieses Gespräch, in dem er ''\emph{Eure fürstliche Gnaden auf etliche hohe Weise und nützliche Fragen nach vermögen geantwortet}'' habe \cite{Schoenfeldt1558}. 

Ohne Zweifel muss dieses Gespräch einen hohen Einfluss auf die weitere Beziehung Schönfeldts zur Familie der Landgrafen gehabt haben. Denn bald darauf lernte Schönfeldt auch Philipp, den Vater Wilhelms, und die Geschwister Wilhelms kennen. Nicht nur avancierte er zum Leibarzt der Familie \cite{Saloch2006}, sondern er diskutierte mit Wilhelm und seinen Brüdern auch astronomische Phänomene wie z.B. Kometen und Mondfinsternisse (siehe \cite{Wilhelm1586} und \cite{Wilhelm1580}).

\section{Astronomische Schriften Schönfeldts}

Die Erstellung von Prognostica gehörte sicher nicht zu den Pflichtaufgaben eines Mathematikprofessors im 16. Jahrhundert, denn nur ein sehr geringer Teil der nach der Erfindung der Buchdruckerkunst aufkommenden Prognostica und Schreibkalender stammten von Hochschullehrern \cite{Kremer2006}. Schönfeldt lernte in Wittenberg die astronomischen Grundlagen und deren Deutungen nach den Schriften der alten Philosophen und Astronomen kennen und sah es wohl als seine Aufgabe an, regelmäßig Prognostica für das kommende Kalenderjahr zu verfassen. 

So hatte er nach Dienstantritt als Hochschullehrer im Frühjahr 1557 sich auch gleich an die Abfassung seines ersten Prognosticon Astrologicum gesetzt. In seiner Kometenschrift von 1558 \cite{Schoenfeldt1558} erwähnt er dieses Prognosticon, ''\emph{das ich Eurer Fürstlichen Gnaden zur unterthenigen verehrung auf diss 1558. jar gestellt}'' habe. Die Kometenschrift wurde nach Beobachtung des Kometen im August 1558 verfasst. In dieser Schrift erwähnt Schönfeldt nicht nur den Besuch bei Wilhelm IV.\ vor eineinhalb Jahren, sondern auch, dass in diesem Prognosticon, welches er ''\emph{im vergangenen 1557. jar / ampts halben gestellt}'' habe ''\emph{deutlich von dieses erschienen Cometen zeit/ und auch ursachen gehandelt ist worden}''.
Es ist daher davon auszugehen, dass das erste Prognosticon von Victorinus Schönfeldt für das Jahr 1558 erschienen ist. 

Bis zu seinem Tode im Jahr 1591 hatte Schönfeldt dann insgesamt 35 Prognostica verfasst. Das letzte Prognosticon war fertig gestellt, bevor er am 13. Juni 1591 verstarb. Es wurde von seinem Sohn Burckhard am 23.\ Juli 1591 herausgegeben. 20 Prognostica liegen dem Autor in digitaler Form vor, drei weitere sind bekannt, aber noch nicht digitalisiert, der Rest gilt momentan als noch nicht aufgefunden oder verschollen.

Neben den ausführlichen Prognostica waren im 16.\ und 17.\ Jh.\ auch Almanache und Schreibkalender sehr weit verbreitet, die allerdings keine Erläuterungen und Erklärungen sondern nur Symbole als Hinweise von Deutungen der Stellungen von Sonne, Mond und Planeten enthielten. Schönfeldt erwähnt in seinen Prognostica häufiger auch seine Almanache, doppelseitige Kalender für jeden Monat des Jahres. Es ist davon auszugehen, dass diese auch für jedes Jahr ab 1558 erstellt wurden. Auch Schreibkalender sind von Victorinus Schönfeldt überliefert, der früheste für das Jahr 1564. Bei diesen Kalendern fand sich auf der linken Seite ein Halbmonat eines Jahres und die rechte Seite war für persönliche Eintragungen freigehalten.
Mehr zu den Almanachen und Schreibkalendern von Schönfeldt ist im Handbuch der Kalendermacher von Klaus-Dieter Herbst zu finden \cite{Herbst}.

Zu besonderen astronomischen Anlässen oder für ihn wichtigen Themen verfasste Schönfeldt themenbezogene Schriften. In den Prognostica werden diese erwähnt, es wird Bezug darauf genommen. Diese Schriften sind also als inhaltliche Vertiefungen der Prognostica zu werten. Bekannte und erwähnte Schriften sind 
\begin{itemize}
	\item Kometenschrift von 1558 \cite{Schoenfeldt1558}
	\item Verzeichnis der Sonnen- und Mondfinsternisse ab 1576 \cite{Schoenfeldt1575}
	\item Kometenschrift von 1577, wird in Prognosticon von 1579 erwähnt, bisher ist kein erhaltenes Exemplar bekannt
	\item Planetenschrift von 1590 \cite{Schoenfeldt1590b}
	\item Heller Stern Arcturus von 1591 \cite{Schoenfeldt1591b}
	\item Planetenschrift von 1591 \cite{Schoenfeldt1591c}
\end{itemize}

\section{Schönfeldt als Astronom und Astrologe}

Um die Prognostica besser verstehen und einordnen zu können, ist es nützlich, zunächst einen Blick auf die Hintergründe zu werfen, auf die Tradition, in der Schönfeldt ausgebildet wurde und die ihn prägte. 

Schönfeldts Ausbildung an der Universität Wittenberg fiel in den Beginn einer neuen Epoche. Bisher wurden für die Berechnung der Gestirnspositionen die Alfonsinischen Tafeln verwendet, die im 13. Jh. auf Anordnung von Alfons X.\ von Kastilien und Le\'{o}n auf Grundlage des ptolemaischen Modells erstellt wurden.  Im Jahr 1543 veröffentlichte Nicolaus Copernicus die Grundlagen des heliozentrischen Modells für unser Sonnensystem. Erasmus Reinhold (1511 -- 1553) war seit 1536 Professor für höhere Mathematik in Wittenberg und hatte sich mit der Weiterentwicklung des copernicanischen Modells beschäftigt. 1551 gab er die sogenannten Prutenischen Tafeln heraus, die zur Berechnung der Positionen der bekannten Himmelskörper des Planetensystems dienten. Er nutze nicht nur aktuelle Ausgangsdaten der Himmelskörper sondern auch das heliozentrische Modell von Copernicus als Grundlage.

Da das copernicanische Modell keine naturwissenschaftliche Begründung für den Wechsel vom geozentrischen zum heliozentrischen Weltbild lieferte, standen Schönfeldts Lehrer Peucer, Melanchton und Reinhold in der Interpretation der Bewegungen der Himmelskörper noch ganz in der alten Tradition des Ptolemaios. In der Wittenbergschen Interpretation des copernicanischen Weltbildes wurde das copernicanische Modell für die Berechnung der Gestirnspositionen akzeptiert, aber kein Urteil über die Richtigkeit des ptolemaischen oder copernicanischen Modells abgegeben \cite{Kremer2006}.

In den frühen Prognostica erläutert Schönfeldt seinen Standpunkt: 
\begin{itemize}
	\item Schönfeldt schreibt von ''\emph{der hohen und elden Kunst der Astrononey / one welcher gewissen grundt gar keine praedictiones oder observationes effectum können ersuchet werden}'' und fordert, 
	die ''\emph{Mathematicii}'' sollen ''\emph{mit größerem Ernst/ denn es geschicht/ die Apparitiones coelestes, als denn sein die loca Stellarum secundum longun \& latum, Ortus \& decubitus, congressus \& alias interse Syzygias, distantias \& configurationes an hellem und klarem Himmel observieren}'' \cite{Schoenfeldt1561}.
	\item Und ebenso schreibt er von ''\emph{der löblichen und nützlichen Kunst/ der Astrologia, welche die Bedeutung und Wirkung der Gestirns über diese undere Welt, welche sich nach mancherley vermischung und zuthuung der vier Element nach formieret und verendert umbgeben}'' \cite{Schoenfeldt1561}.
\end{itemize}

\subsection{Schönfeldt als Astronom}
\label{sec_schoenfeldt_astronom}

Schönfeldt ärgerte sich immer wieder über Verfasser von Prognostica, die nicht selber beobachteten und so z.B. nicht feststellten, dass der Merkur überhaupt nicht da aufzufinden sei, wo er nach den alten Tabellen stehen sollte \cite{Schoenfeldt1561}. Er selbst beobachtete oft mit seinem Torquetum auf dem Schlossberg hinter dem Schloss der Landgrafen in Marburg, wie ein Ausschnitt der Beobachtung des Kometen von 1558 zeigt: ''\emph{Nach diser observation bin	
ich als bald widerum herein auff den Schloßberg gangen/ auff welchem ich meine Instrumenta zuvormals etliche tag hette zugestelt/ und alda des Cometen gewissen lauff an der lenge / breyte von der Ecliptica/ und der höhe von der Erden/ neben anderen Stücken mehr/ welche zu der rechnung und abmalung dinstlichen/ ersuchet}'' \cite{Schoenfeldt1558}.

Schönfeldt hatte regelmäßig die Breite von Marburg gemessen, die er auch ab 1569 in den Prognostica aufführt: ''\emph{Latitudinem Marpurgensem / Welche sich über 50 grad 50 minuten erstrecket / wie sie allhier fleissig observieret worden}'' \cite{Schoenfeldt1570}. 

\begin{figure}[H]
		\begin{center}
		\includegraphics[width=0.6\linewidth]{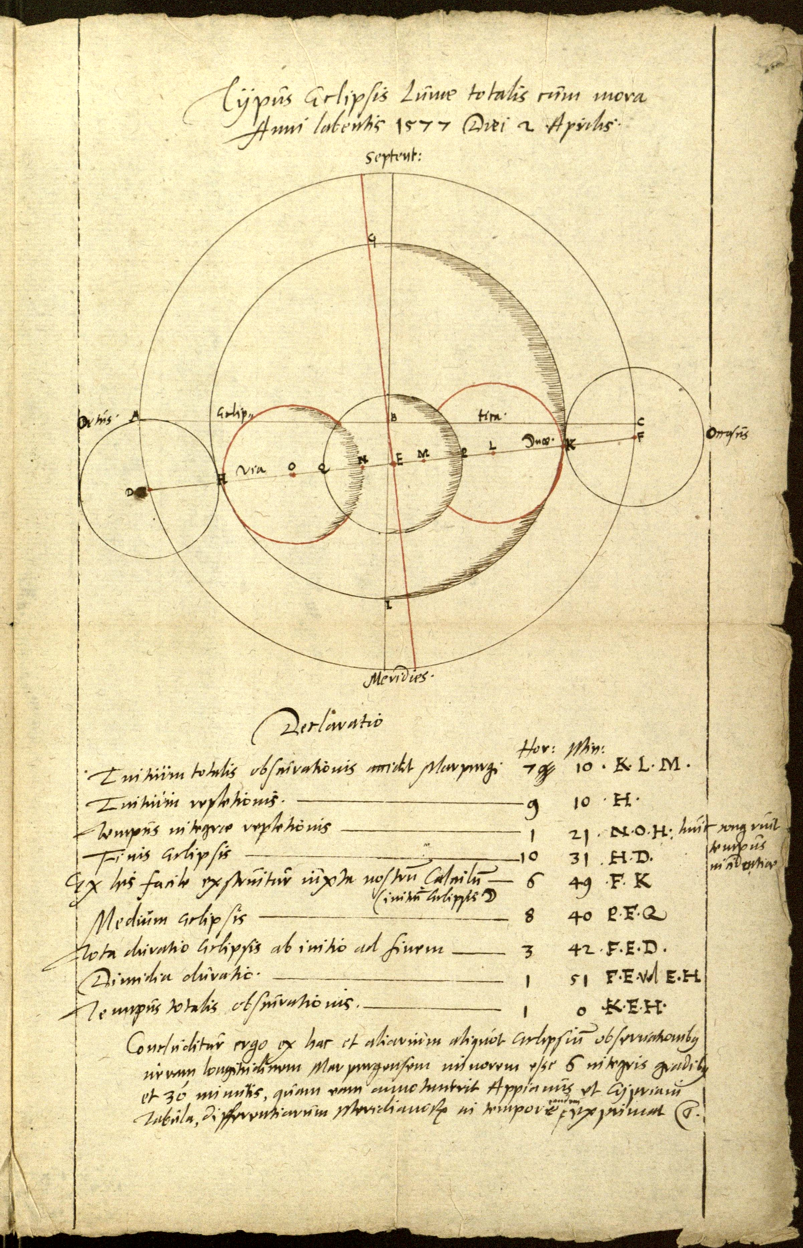}
		\caption{Beobachtungsblatt zur Mondfinsternis vom 2.\ April 1577 von V.\ Schönfeldt in Marburg, Anlage zu einem Brief von Schönfeldt an Wilhelm IV.\ \cite{Schoenfeldt1577b}. Scan: Hessisches Staatsarchiv Marburg}
		\label{fig_mofi_1577}		
	\end{center}
\end{figure}

Im April 1577 schickte Schönfeldt an Landgraf Wilhelm IV.\ einen Brief mit der Beobachtung der totalen Mondfinsternis vom 2.\ April 1577 (Abb.\ \ref{fig_mofi_1577}).

Ein weiteres Beispiel für Schönfeldts astronomische Messungen sind die Beobachtungen im Sternbild des Bootes, vernehmlich am Stern Arktur \cite{Schoenfeldt1591b}. Er listete dort eigene Positionsmessungen des Arktur sowie dessen Distanzen zu anderen hellen Sternen auf, erläuterte die Sterne des Sternbild Bootes und schlug vor, Auf- und Untergang von Arktur relativ zur Stellung der Sonne als ein für jedermann leicht feststellbares Zeichen für die Jahreszeiten zu verwenden.

\subsection{Astrologie im 16. Jahrhundert}
\label{sec_schoenfeldt_astrologie}

Die Verbindung der Astronomie und Astrologie war in der damaligen Zeit naheliegend. Es gab noch keine Begründung für die Bahnen der Himmelskörper unseres Planetensystems, bis zur Beschreibung der Gravitation sollte es noch ca. einhundert Jahre dauern. Also lieferte die Astrologie die Erklärung der Himmelsphänomene.

Schönfeldt hatte einen urchristlichen Ansatz, der alle seine Prognostica durchzieht und den sein Sohn im Vorwort des letzten Prognosticons treffend beschreibt: ''\emph{Es wird niemand, der verstendig/ und dieser Kunst recht erfahren/ auskratzen oder austilgen/ das Oraculum/ das Gott selbst lassen erschallen und verlauten baldt am anfang seiner Prophetischen Bücher / da er also spricht:... }''. Und dann zitiert er Genesis 1, Verse 14 und 15 ''\emph{Es werden Liechter an der Feste des Himmels/ die da scheiden Tag und Nacht/ und geben Zeichen/ Zeiten/ Tage und Jahre/ und seien Liechter an der Feste des Himmels/ die da scheinen auff Erden, etc.}'' \cite{Schoenfeldt1591a}.

In seinem Prognosticon für das Jahr 1561 führte Schönfeldt sehr detailliert aus, welche Zeichen die Lichter an der Feste des Himmels geben \cite{Schoenfeldt1560}: ''\emph{was für seltsam und fast ungleubliche dinge / sich in unvernünftigen Thieren / Edlen Steinen und Gewechsen der erden / sehen und vermercken lassen / welche fürnemlich nach dem lauff und bewegung des Himels / und sonderlich der irrende Stern/ welche man Planeten nennet/ zu geschehen pflegen}''. Es folgt eine längere Auflistungen von Beispielen für Veränderungen in Aussehen und Verhalten im Rhythmus der Sonne und des Mondes aus der Pflanzen- und Tierwelt. Im Prognosticon für das Jahr 1563 \cite{Schoenfeldt1562} erläuterte Schönfeldt die Wechselwirkung zwischen den Planeten und den Geschehnissen auf der Erde, die durch das Licht der Planeten vermittelt werden sollte. 
In dieser Abhandlung wird die damalige Vorstellung von Lichtstrahlen, Farben der Planeten und der Kristalle und anderer Dinge auf der Erde diskutiert.

Um in den Vorhersagen einen Ortsbezug herzustellen, gab es eine Zuordnung von Ländern und Städten zu den Sternzeichen. Im Anhang zum Prognosticon für das Jahr 1578 ist ein ''\emph{Register der Stedt und Lande/ so under den himmlischen Zeichen gelegen sind}'' aufgeführt, es sind drei von vier Quadrangel abgedruckt \cite{Schoenfeldt1578}.

\begin{figure}[H]
		\begin{center}
		\includegraphics[width=0.6\linewidth]{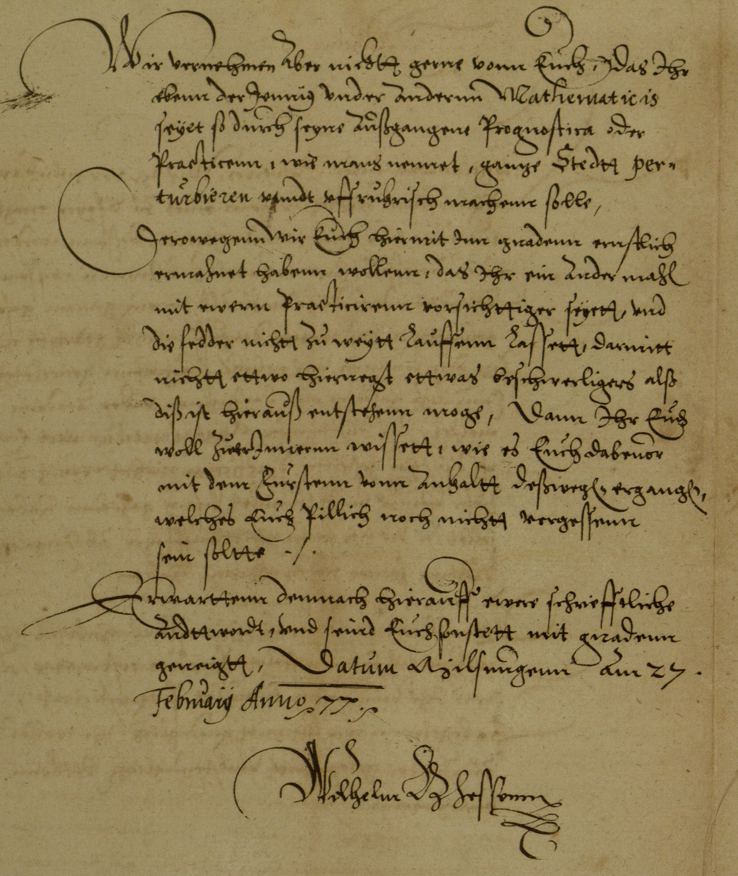}
		\caption{Auszug aus dem Brief von Wilhelm IV.\ vom 27. Februar 1577 an Schönfeldt \cite{Wilhelm1577}. Scan: Hessisches Staatsarchiv Marburg}
		\label{fig_brief_w_iv_1577}		
	\end{center}
\end{figure}

Die Bewahrheitung astrologischer Vorhersagen war auch im Mittelalter eher zufällig, außer es handelte sich um allgemeine Aussagen, die in der Regel zutrafen, wie z.B.\ dass die Planeten im Winter ''\emph{viel Mittagige winde erregen/ nicht ohne sonderliche Kelte und Schnee}'' \cite{Schoenfeldt1562}.
Auch hatten Vorhersagen unter Umständen eine psychologische Wirkung, die dann genau die Dinge hervorriefen, die prognostiziert waren. Schönfeldt kündigte die Wirkung der ''\emph{großen Finsternis des Monden}'' vom 7. Oktober 1576 für das Frühjahr 1577 mit großem Tumulten für die Obrigkeit an: ''\emph{Ihr viel werden überfallen / belegert / eingenommen/ uberschetzt und geplündert werden ... werden viel ernewerung und reformationes leiden und ertragen müssen}'' \cite{Schoenfeldt1577a}. Und dann wies Schönfeldt die Menschen in Speyer auf einen früheren Aufstand nach einer Mondfinsternis im Oktober 1512 hin \cite{Schoenfeldt1577a}. Diese Anmerkung muss im Februar 1577 erneut einen Aufstand in Speyer provoziert haben, für den Schönfeldt von seinem Landgrafen deutlich gerügt wird: ''\emph{Dieweill wir dann gernn ein wissens haben mochtten, wohinn sich Ewer gesteltte Practica erstreckett, auch was Ihr vonn diesem tumultt prognosticiret, so begehrenn wir hiermitt gnediglich Ihr wollet uns zu ehister gelegenheitt umbstendiglich daruonn berichtenn. Wir vernehmen aber nicht gerne vonn Euch, das ihr ebenn derjenig under anderenn Mathematicis seyet so durch seyne außgangene Prognostica oder Practicenn, wie mans nennet, ganze Stedt perturbieren unddt uffrührisch machenn solle}'' (\cite{Wilhelm1577} und Abb.\ \ref{fig_brief_w_iv_1577}).

Als Antwort und zur Beruhigung seines Landesherren schickte Schönfeldt seine Beobachtung der Mondfinsternis vom 2.\ April 1577 an Wilhelm IV.\ (siehe Kapitel \ref{sec_schoenfeldt_astronom}).

\section{Die Prognostica}
Es ist davon auszugehen, dass Schönfeldt insgesamt 35 Prognostica für die Jahre von 1558 bis 1592 verfasst hat. Das grobe Schema ist in allen Prognostica recht ähnlich. Maßgebliche Unterschiede ergeben sich vermutlich aus der Zeit, die Schönfeldt zum Abfassen dieser Jahresbücher zur Verfügung hatte. Die hier aufgeführten Erläuterungen beruhen auf einer Auswertung der Prognostica für die Jahre 1561 -- 1564, 1567 und 1568, 1570 -- 1579 und 1588 -- 1592.

\subsection{Aufbau}
\begin{itemize}
	\item Titelblatt mit Grafik
	
	Das Titelblatt enthält immer eine große dominierende Grafik, oft Darstellungen von Krieg, Not, Unwetter mit partiellen Finsternissen oder Planeten am Himmel. Selten sind die Grafiken aus heutiger Sicht freundlich oder ansprechend gestaltet. Einige Male wurden Teile oder auch komplette Grafiken im Folgejahr nochmal verwendet, wobei Inschriften verändert wurden.
	
	\item Kapitel mit astronomischen Erläuterungen
	\begin{itemize}
		\item Widmung mit langer Vorrede
		
		Nur in den frühen Prognostica bis zu dem für das Jahr 1564 und den späten ab dem Jahr 1589 ist eine Widmung mit einer langen Vorrede eingefügt.
		
		Die Widmungen sind an die Landgrafen, hohe Würdenträger der Stadt Marburg, einmal auch an den Bürgermeister von Kassel und im letzten Prognosticon von Schönfeldts Sohn Burckhard an den Bürgermeister der Geburtsstadt Schönfeldts, Bauzen, gerichtet. In den Vorreden erläutert Schönfeld viele astronomische Zusammenhänge, deren wichtigste in den folgenden Kapiteln kurz aufgezeigt werden, sowie auch astrologische Hintergründe (siehe Kapitel \ref{sec_schoenfeldt_astrologie}).
		
		Vermutlich hat Schönfeldt wegen zunehmender Verpflichtungen sowohl als Mediziner als auch in der Verwaltung der Universität auf die Ausführungen in der Vorrede verzichtet. In den späten Prognostica deutet er in den Vorreden an, dass seine Jahrbücher gerne gelesen und beachtet wurden. Man kann daraus erahnen, dass er gebeten wurde, wieder ausführlichere Vorworte einzufügen.
		
		\item Kapitel über die vier Jahreszeiten
		
		In diesen Kapiteln werden die berechneten astronomischen Wechsel der Jahreszeiten genau angegeben, öfters auch nach Berechnungen auf Grundlage mehrerer verschiedener Tafeln (Kapitel \ref{sec_astronomische_tafeln}). Außerdem werden die Sichtbarkeit und der Lauf der Planeten wiedergegeben.
		
		\item Kapitel über die Finsternisse
		
		Schönfeldt listet die Daten der im Jahr des Prognosticons auftretenden sichtbaren Sonnen- und Mondfinsternisse auf.

	\end{itemize}
	
	\item Kapitel mit astrologischen Deutungen
	
	Die Kapitel mit den astrologischen Deutungen behandeln Krankheiten, Krieg und Ackerbau und tragen Überschriften wie ''\emph{Von Pestilenz und anderen Leibs Schwachheiten }'', ''\emph{Von Krieg, Uneinigkeit und Unfriede }'' und ''\emph{von Früchten und Gewächsen auf Erden}''.
	
	\item Abschluss
	
	In den allermeisten Prognostica endete Schönfeldt mit einem Segenswunsch und einem Bibelvers, gelegentlich auch einem kurzen Gedicht in Latein statt des Bibelverses.

\end{itemize}

\subsection{Astronomische Tafeln}
\label{sec_astronomische_tafeln}

Durch seine Ausbildung in Wittenberg geprägt wusste Schönfeldt natürlich um die neuen Tabellen von Reinhold, die auf der Grundlage des copernicanischen Weltbildes beruhen. Seine Verehrung für Copernicus hielt er nicht zurück: im Prognosticon für 1562 schreibt er ''\emph{Es wird aber die Sonne / wie es am gründlichsten des wolerfarnen Mathematici Nicolai Copernici Rechnung aufweist / eben diesen himlischen Punkten des Principji Dodecatemorij Cpariconri ... (am) ... erreichen }'' \cite{Schoenfeldt1561}, und im Prognosticon für 1592 lobt er ''\emph{Copernicus / als ein vornehmer und vleissiger Astronomus / ... }''
\cite{Schoenfeldt1591a}.

Ab dem Prognosticon für 1562 verwendete Schönfeldt für seine Rechnungen das copernicanische Modell für seine Berechnungen.
Dies lässt sehr darauf schließen, dass auch die früheren Prognostica auf dieser Grundlage erstellt wurden. Explizit angemerkt hat Schönfeldt dies im Prognosticon für 1561 (das früheste dem Autor vorliegende Prognosticon) jedoch nicht.

\begin{figure}[H]
	\begin{center}
		\includegraphics[width=0.6\linewidth]{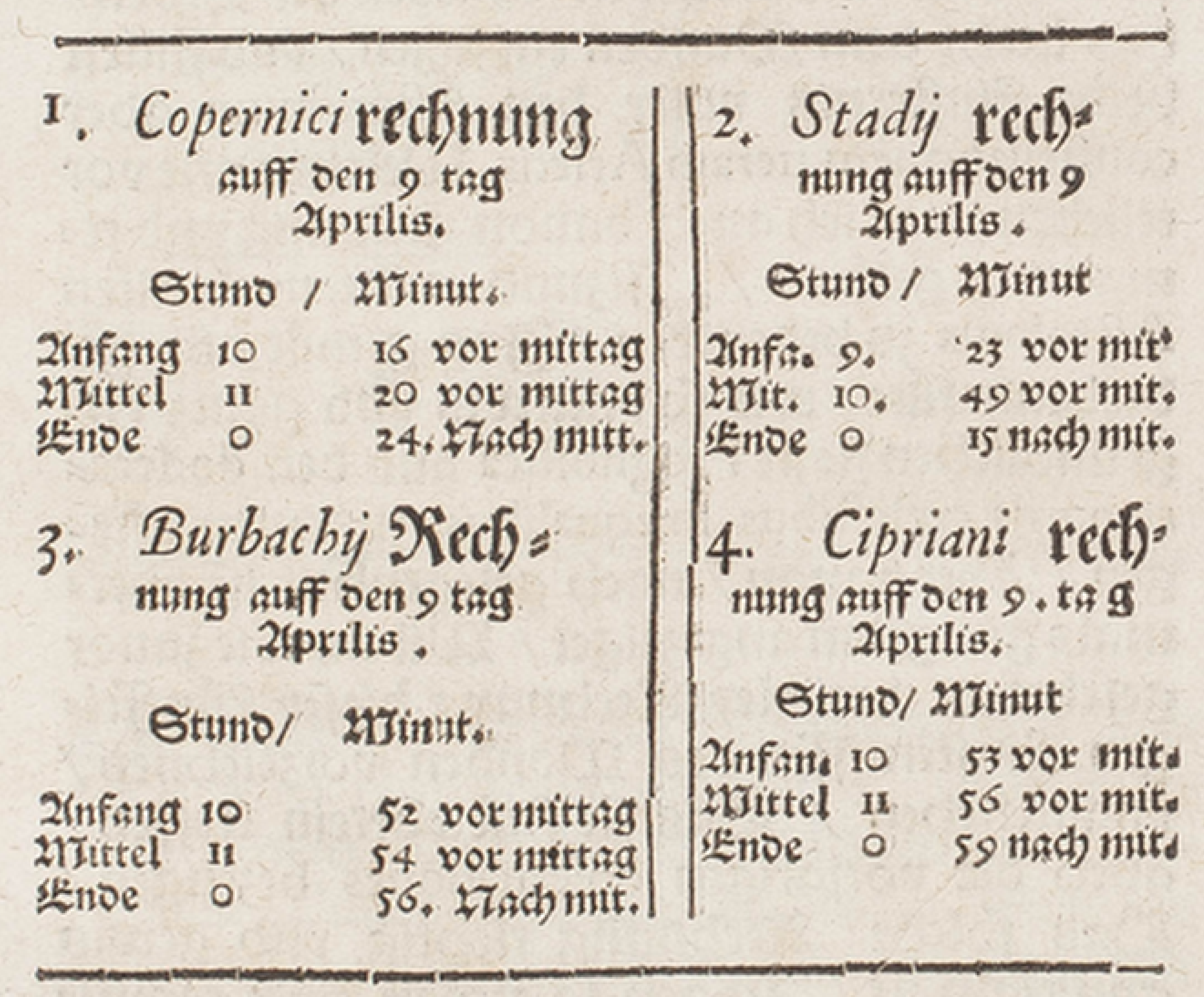}
		\caption{Berechnung der Sichtbarkeit der Sonnenfinsternis vom 9. April 1567 von Marburg aus \cite{Schoenfeldt1567} nach vier verschiedenen Tafeln und Modellen. Scan: Herzog August Bibliothek Wolfenbüttel.}
		\label{fig_sofi_1567}
	\end{center}
\end{figure}

Im Prognosticon für 1567 taucht zum ersten Mal ein Vergleich von vier Vorhersagen für eine Sonnenfinsternis auf, die mit verschiedenen Tafeln und Modellen berechnet wurden (Abb. \ref{fig_sofi_1567}). Neben der Berechnung nach Copernicus (Nr.\ 1 in Abb. \ref{fig_sofi_1567}) verwendete Schönfeldt auch die von Johannes Stadius (1527 -- 1579) aus den Prutenischen Tafeln abgeleiteten Ephemeridentafeln (Nr.\ 2 in Abb. \ref{fig_sofi_1567}), sowie die auf den Alfonsinischen Tafeln beruhenden Tabellen von Georg von Peuerbach (1423 -- 1461) (Nr.\ 3 in Abb. \ref{fig_sofi_1567}) und von Cyprian Leowitz (1524 -- 1574)  (Nr.\ 4 in Abb. \ref{fig_sofi_1567}) und schreibt dazu: ''\emph{Damit man endlich erfahre welche tabular oder Rechnung den augenscheinlichen observationibus am nechsten zufallen}'', ganz in der Wittenberger Tradition, keine Festlegung zu treffen.

Diese Vergleiche tauchen in den Prognostica hin und wieder auf. In den späten Prognostica werden sie ergänzt oder auch ersetzt durch neuere Tabellen. So nutzte er für das Prognosticon für 1591 die 2.\ Auflage der Prutenischen Tafeln, die von Michael Maestlin (1550 -- 1631) 1571 herausgegeben wurde, sowie Tafeln des ''\emph{New fundamento}''. Eine Erläuterung, was er damit meinte, ist im Prognosticon für 1591 nicht zu finden. Dafür taucht im Prognosticon für 1592 der Name Tycho Brahe (1546 -- 1601) auf: ''\emph{die Neoterici/ sonderlich Tycho a Brahe}'' und vermutlich ist der ''\emph{Neotericus}'' des Prognosticons von 1592 auch die Urheber des ''\emph{New fundamento}'' im Jahr zuvor.  Das heißt, Schönfeldt verwendete in seinem letzten Prognosticon das Tychonische System, welches Brahe in 1588 zum ersten Mal publizierte.

\subsection{Sonne, Mond und Planeten -- Himmelskörper der Ekliptik}

Schönfeldt wandte seine astronomischen Rechnung zur Positionsbestimmung der bekannten Himmelskörper unseres Sonnensystems an.

Als erstes ist natürlich die Sonne zu nennen, Schönfeldt berechnete den Beginn der vier Jahreszeiten, was ja schon den allermeisten Titeln der Prognostica angedeutet ist. Ab dem Prognosticon 1567 gibt er in Tabellenform für jede Jahreszeit den Eintritt, Verweildauer und Austritt der Sonne in und aus dem jeweiligen Sternzeichen an.

Der Lauf des Mondes und der Planeten sowie die Mondphasen werden in Worten beschrieben, es werden keine Daten in Tabellenform angegeben.
Annäherungen der Planeten untereinander sowie des Mondes zu den Planeten werden taggenau aufgeführt.

Besondere Ereignisse -- auch für die Vorhersagen -- sind die Finsternisse. Schönfeldt berechnete für einen Beobachter in Marburg den Zeitpunkt und den Grad der maximalen Bedeckung sowie die Dauer der Sonnen- und Mondfinsternisse. 

Bei seinen Erläuterungen im Vorwort zum Prognosticon für 1562 geht Schönfeldt auch auf die Verschiebung des Frühlingspunktes ein: ''\emph{das fast alle hundert jar die sichtbaren Stern des himels  sich von den gradibus Dodecatemoriorum equalium auf einen gradum/ doch wol unmerklichen absondern }'' \cite{Schoenfeldt1561}. Als Beispiel nannte Schönfeldt den Beginn der Hundstage (den heliakischen Aufgang des Sirius), der 432 v. Chr. zur Zeit des Hippokrates auf den 13. Juli, im Jahr 65 n. Chr. auf den 15. Juli und 1562 auf den 17. Juli fiel \cite{Schoenfeldt1561}.

\subsection{Kalenderrechnung}

Schönfeldt verfasste alle Prognostica nach dem alten Julianischen Kalender. Die Kalenderreform von 1582 wurde in Hessen erst im Jahr 1700 umgesetzt \cite{Hamel2002}.

Im Prognosticon von 1576 erläuterte Schönfeldt die Entstehung des julianischen Kalenders durch den Auftrag Julius Caesars an Sosigenes aus Alexandria. Der Schalttag im alten Kalender war im Gegensatz zum heutigen Schalttag der 5. letzte Tag im Februar. In Schaltjahren gab es zwei Tage, die man 24. Februar nannte \cite{Schoenfeldt1576}.

Die Länge des tropischen Jahres, die den Kalender letztendlich bestimmt, berechnete Schönfeldt nach dem copernicanischen Weltbild für die Jahre 1562, 1570 -- 1573. Z.B. für 1573 erhielt er 356 Tage, 5 Stunden, 55 Minuten, 22 Sekunden und 27 Tertien (1/60 Sekunde), dies ergibt 365.246 78 Tage, die anderen Werte unterscheiden sich  nicht um mehr als 0.000 05 Tage. Im Prognosticon für 1562 stellte Schönfeldt fest, dass dieser Wert deutlich größer als derjenige ist, der sich aus den Alfonsinischen Tafeln berechnen ließ, etwa 6 Minuten. Damit verschoben sich die Ergebnisse von Rechnungen mit diesen beiden Systemen im Laufe der Jahre immer mehr. Schönfeldt nutzte für seine Berechnung des tropischen Jahres jedoch ausschließlich das copernicanische System, so dass man zu dem Schluss kommen kann, dass er diesem Weltbild mehr Vertrauen entgegen brachte. Anzumerken ist, dass das Resultat aus den Alfonsinischen Tafeln dichter beim wahren Wert von 365.242 196 Tagen lag als das Ergebnis nach den copernicanischen Rechnungen.

Schon in seinem Prognosticon für 1562 ging Schönfeldt auf die aktuellen Probleme der Kalenderrechnung ein. Der julianische Kalender gab die mittlere Jahreslänge mit 364.250 Tagen zu groß an, so dass es im 16. Jahrhundert zu einer Verschiebung des gesamten Kalenders und vor allem der christlichen Feiertage von 10 Tagen kam. Schönfeldt hielt sich aber diplomatisch aus dieser Diskussion heraus: ''\emph{aber ich gedencke/ wie gesagt/ keinem kein gewis maß noch ziel fürzustellen/ es gehet Könige/ Fürsten und recht fleißige Theologos an/ welchen sonst bey den alten fürnemlichen die Kirchenrechnung befohlen/ und vertrawet ist worden}'' \cite{Schoenfeldt1561}.

Im Prognosticon für 1589 formulierte Schönfeldt recht deutlich seine Ablehnung der Kalenderreform: ''\emph{Es hat wol für kurzen Jaren Bapst Gregorius der dreyzehende sich gelüsten lassen/ und alleine aus lauterem Ehrzeiz/ das alle gemeine breuchliche Allmanach zu bessern und zu endern unterstanden. Was aber diß sein vornemen für ein Fundament habe/ ist vor der zeit durch viel verstendige Mathematicos ausfürlichen angezeigt/ und als untüchtig verworfen worden/ und wird noch mehr desselbigen jrrthumb gnungsam bescheinet werden/ unangesehen/ das newlicher zeit Antonius Possevinus, Ein Jesuiter des obbenandten Babst Correction unnd reformation des Calendarij/ sich mit schlechtem beweis zu bestedigen unternomen/ und noch sich in diesem bemühet/ wird aber zu dieser zeit geringen lob erlangen/ und seinen lohn erwarten müssen.}''

\subsection{Kometen}

Neben den Finsternissen waren die Erscheinungen der Kometen astronomische Phänomene, die von jedermann wahrgenommen werden konnten. Mindestens zwei wichtige Dinge prägten die Kometen: die Unberechenbarkeit der Vorhersage ihres Auftretens und die Natur des Schweifs.

\begin{figure}[H]
	\begin{center}
		\includegraphics[width=0.6\linewidth]{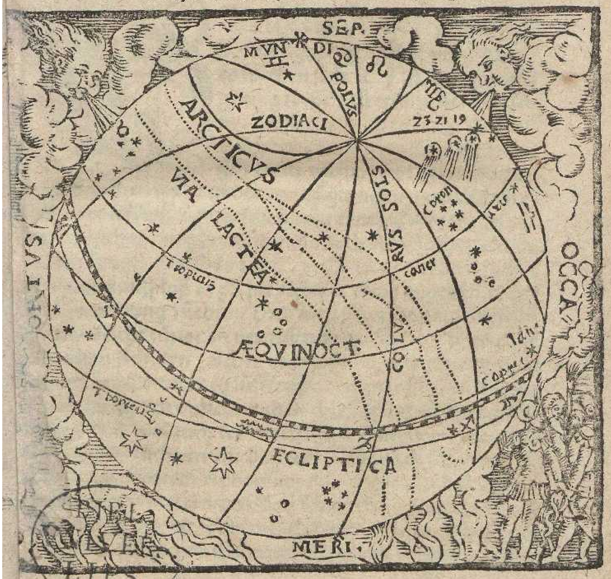}
		\caption{Skizze mit den Beobachtungen des Kometen von 1558 von Marburg aus \cite{Schoenfeldt1560}. Eingetragen sind die Beobachtungen vom 19., 21.\ und 23.\ August. Scan:  Universitätsbibliothek Leipzig.}
		\label{fig_komet_1558}
	\end{center}
\end{figure}

Schönfeldt hatte sicherlich alle Kometenerscheinungen beobachtet, die während seiner Lebenszeit auftraten. Überliefert sind, dass er den Kometen von 1556 in Wittenberg beobachtet hatte \cite{Bechstein1875}, die ausführlichen Beobachtungen des Kometen von 1557 (Abb.\ \ref{fig_komet_1558}), die in seiner Kometenschrift \cite{Schoenfeldt1558} publiziert wurden, die Beobachtung des Kometen von 1577, die in einem Brief von Wilhelm IV.\ an Schönfeldt \cite{Wilhelm1586} und in mehreren Verweisen Schönfeldts in seinem Prognosticon für 1579 \cite{Schoenfeldt1579} und die Beobachtungen des Kometen von 1580, über die ein Briefaustausch mit den drei Landgrafen Wilhelm IV.\, Ludwig IV.\ und Georg I.\ vorliegt \cite{Wilhelm1580}. 

Der Schweif der Kometen konnte nach der damaligen Vorstellung nur aus irdischen Gasen bestehen, giftige brennende Gase, wie Schönfeldt in seiner Kometenschrift ausführt, die durch die Kraft der Sonne und der Planeten aus der Erde ausgelöst und in die Höhe gezogen werden \cite{Schoenfeldt1558}. 

Schönfeldt erläuterte, dass Kometen aus zwei Teilen bestehen, aus einem Kern und einem Schweif. Bekannt und verwunderlich war, dass der Schweif keineswegs immer zur Erde hinunter zeigt. Stattdessen, so zitiert Schönfeldt Girolamo Fracastore (1476 -- 1553), sollte sich ''\emph{des Cometen Schanz \`{e} directo in oppositam Solis partem richten}'', der dieses Phänomen durch genaue Beobachtungen des Halleyschen Kometen 1531 herausfand und beschrieb.

Interessant ist auch, dass Schönfeldt in der Vorstellung, dass Kometen sich zwischen Erde und Mond bewegen müssten, da sie ja von der Erde stammten, eine Parallaxenmessung, also eine Entfernungsbestimmung, vorschlug: ''\emph{welcher sich der Astronomiae unterstehet \ldots der gründlichen observation und rechnung sich fleisset/ auffs aller höchste und freundlichste gebeten und ermahnet haben/ wölle sich mit gerechten Instrumenten und so es möglichen parallactici/ rüsten und gefasst machen }'' \cite{Schoenfeldt1558}. Damit befand sich Schönfeldt in dieser Frage auf aktuellem Stand der Forschung, denn sowohl in der Sternwarte Wilhelm IV.\ in Kassel als auch in Brahes Sternwarte auf der Insel Hven untersuchte man Ende des 16. Jahrhunderts genau diese Frage.

\section{Popularisierung der Astronomie}

Schönfeldts Schriftenreihe \emph{Prognosticon Astrologicum} enthält sowohl astronomische Daten wie auch deren astrologische Interpretation. Schaut man genauer hin, dann bemerkt man, mit welcher großen Sorgfalt Schönfeldt seinem Anspruch nachkommt, keine unpräzisen astronomischen Fakten zu nennen. Er ist einer der wenigen Verfasser solcher Prognostica, die diese als Hochschullehrer erstellen.

Schönfeldts Jahrbücher ähneln in den astronomischen Inhalten modernen Jahrbüchern deutlich: der Lauf der Sonne durch die Sternbilder mit dem Beginn der Jahreszeiten, Mondphasen und Planetensichtbarkeiten werden aufgelistet. Besondere Konstellationen wie Finsternisse oder Konjunktionen spielen eine wichtige Rolle und Hintergrundinformationen werden in kleineren Aufsätzen hinzugefügt. In diesem Sinne erfüllen Schönfeldts Jahrbücher den gleichen Sinn wie heutige Jahrbücher, den der Popularisierung der Astronomie!

Durch vielseitige Verpflichtungen gebunden reduzierte Schönfeldt seine Jahrbücher zwischen 1564 und 1588. Aber, die Jahrbücher müssen sehr beliebt gewesen sein, wie Schönfeldt z.B.\ im Vorwort des Prognosticons für 1591 erwähnt \cite{Schoenfeldt1591a}, weshalb er sich in den letzten Jahren wieder für ausführlichere Prognostica entschied.

Über die Auflagen seiner Jahrbücher kann man nur spekulieren. Die hohe Vielzahl von Almanachen, Schreibkalendern und Prognostica im 16.\ und 17. Jh.\ legen eine hohe Auflage von einigen Tausend bis Zehntausend Exemplaren nahe \cite{Herbst2016}. Eine Bemerkung von Schönfeldts Sohn im Vorwort des letzten Prognosticons unterstützt diese Einschätzung. Er erwähnt, dass es in der damaligen Türkei keine Verfasser solcher Prognostica gab, weswegen ''\emph{sie alle Jahr viel tausend exemplar Calendarii von den unseren geschrieben/ zu Venedig lassen auffkeuffen/ und in die Türckey hin und wieder verschicken.}'' \cite{Schoenfeldt1591a}. Bemerkenswert ist auch, dass 1595, also vier Jahre nach Schönfeldts Tod, ein Prognosticon eines Autors Johann Victorin Schönfelt aus Marburg in Hessen erschien \cite{Herbst}. Dabei kann es sich nur um eine Fälschung handeln, bei der ein fremder Autor die Bekanntheit von Schönfeldts Prognostica für seinen eigenen wirtschaftlichen Erfolg nutzen wollte.

\section*{Danksagung}

Auf Victorinus Schönfeldt wurde ich anlässlich eines Besuches von Jürgen Hamel bei einem Kolloquium in Marburg hingewiesen. Für die Weitergabe wertvoller Notizen zu den Unterlagen über Schönfeldts Verbindung zu Wilhelm IV.\ im Hessischen Staatsarchiv Marburg bedanke ich mich herzlichst!

Klaus-Dieter Herbst hat mit seinem Projekt des Bibliographischen Handbuchs der Kalendermacher von 1550 bis 1750 die Recherche über Schönfeldt wesentlich gefördert und erleichtert und in persönlichen Gesprächen viele Hinweise gegeben. Auch ihm gebührt ein herzliches Dankeschön!

%\bibliographystyle{plainnat}

% \pagebreak

\end{document}